# Simulated avalanche formation around streamers in an overvolted air gap

Chao Li, Ute Ebert, Willem Hundsdorfer

***Abstract*** - We simulate streamers in air at standard temperature and pressure in a short overvolted gap. The simulation is performed with a 3D hybrid model that traces the single electrons and photons in the low density region, while modeling the streamer interior as a fluid. The photons are followed by a Monte-Carlo procedure, just like the electrons. The first simulation result is present here.

Photo-ionization is recognized as an important ingredient for the propagation of positive streamers in air (but not necessarily in other gases) [1]. But in fluid models for negative streamers, the velocity seems to be modified by photo-ionization only in sufficiently high background fields [2,3]. In this work, photo-ionization is modeled on the particle level [4] in our hybrid model [5,6]. This means that single electrons and photons are followed as particles in the low density region to describe the potential run-away of electrons as well as the density fluctuations arising from the discreteness of the quantum particles.

The photo-ionization is implemented in two steps: first ionizing photons are emitted, and then the photons are absorbed in an ionization event. i) The emission rate of ionizing photons is assumed to be proportional to the number of ionizing electron-neutral collisions. The hybrid model couples a particle model with a fluid model in space, as discussed in [5,6]. In both the particle region and the fluid region, photons are generated by a Monte Carlo process based on the local rate of ionizing electron-neutral collisions. ii) The frequencies of ionizing photons are assumed to be uniformly distributed within 2.06 - 2.93 pHz. The mean free path length of photons is a function of the frequency as given in [7]. The flight distance of the photons is then sampled from their mean free path length as for the electrons in [5,6]. Another random number determines the flight angle of photons under the assumption of isotropic emission; distance and angle determine the location of absorption. When absorption occurs in the particle region, a new electron ion pair is created. When absorption occurs in the fluid region, the local particle densities change correspondingly.

The simulation results are shown in Fig. 1. Two planar electrodes at a distance of 1.17 mm create a background field of -100 kV/cm in air at standard temperature and pressure. The initial condition consists of a Gaussian distribution of 500 electron ion pairs near the cathode. The simulation is carried out on a grid of 256 x 256 x 512 points with cell length $\Delta x=\Delta y=\Delta z=2.3\mu m$ with Neumann boundary conditions in the lateral direction. The time step is $\Delta t=$ 0.3 ps. Fig. 1 shows the bulk of the electrons with a density > $10^{13}/cm^3$ (left), the charge density (middle), and the strength of the electric field in *z*-direction $E_z$ (right) at four different times. At t=0.375 ns, a double headed streamer starts to form. At t=0.4875 ns, an avalanche exceeds the density threshold and appears in the electron bulk plot. At t=0.6 ns, an avalanche (at the right of the primary streamer) propagates across the *y*=0 plane and therefore can be seen in all three plots. At t=0.7125 ns, the space charge layer of the primary streamer as well as the electric field are strongly perturbed. Most photons are absorbed within a few μm, but with a small probability, photons travel further and ionize molecules at a rather long distance; and the electrons liberated by these far traveled photons initiate new avalanches.

In conclusion, by introducing a particle description of the photo-ionization in our hybrid model, we are able to study the density fluctuations induced by single photons.

Manuscript received 1 Dec. 2010.
Chao Li is with Department of Applied Physics, Eindhoven University of Technology, The Netherlands, and Ute Ebert and W. Hundsdorfer are with Centrum Wiskunde & Informatica (CWI), Amsterdam, The Netherlands.
Work was supported by project EWR-science of EUV 10010305 and STW-project 10118.
Publisher Identifier S XXXX-XXXXXXX-X

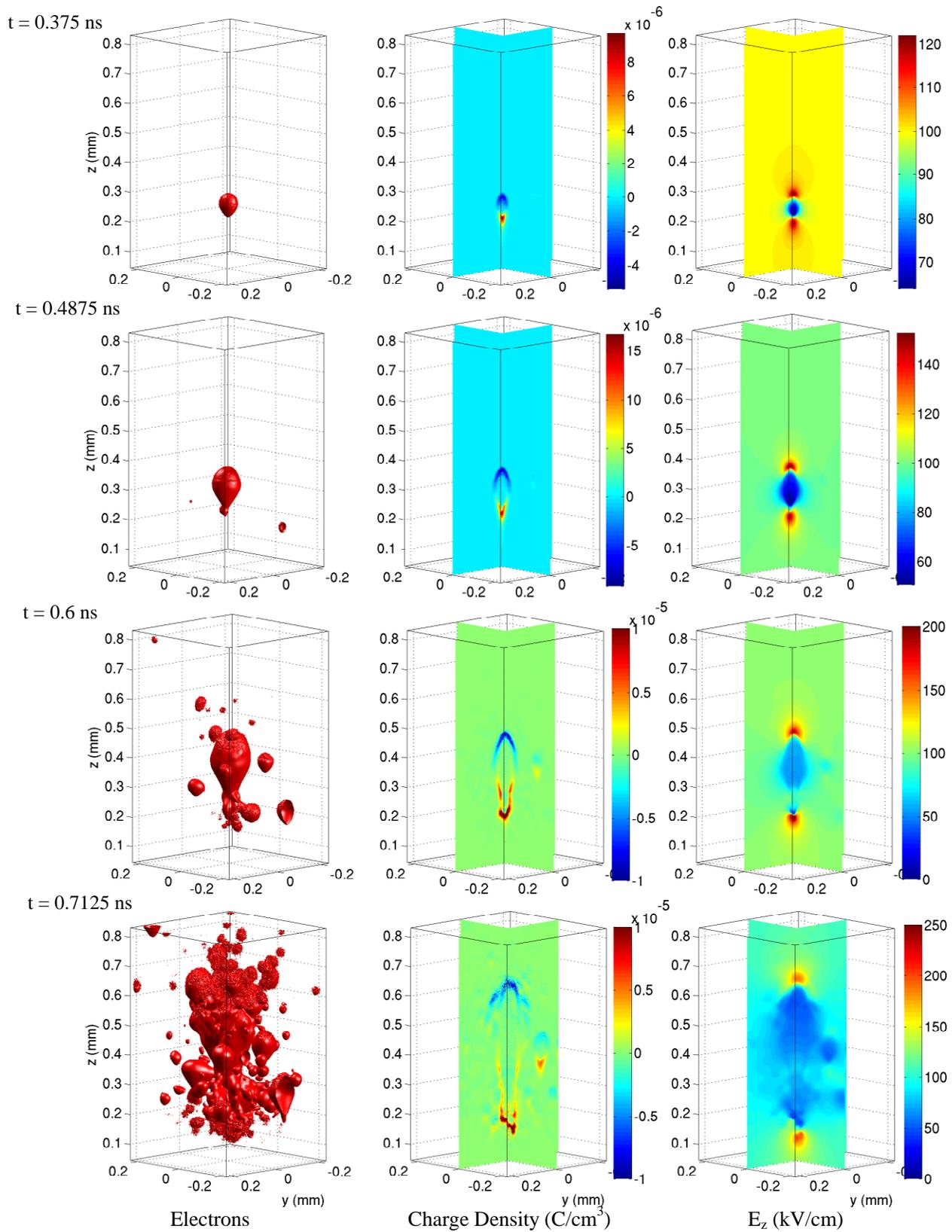

Fig.1 Streamers in air in a background field of -100 kV/cm; electrons drift upwards in this field. First row: t = 0.375 ns, second row: t = 0.4875 ns, third row: t = 0.6 ns, fourth row: t = 0.7125 ns. The columns show from left to right: the bulk of electrons with density > $10^{13}/cm^3$, the charge density ($C/cm^3$), and the electric field $E_z$ (kV/cm). Charge densities and $E_z$ are represented on two orthogonal planes.